\documentclass[sn-apa]{sn-jnl}

\usepackage{graphicx}
\graphicspath{{figures/}}

\usepackage{mdframed}
\usepackage{array}
\newcolumntype{x}[1]{>{\centering\let\newline\\\arraybackslash\hspace{0pt}}p{#1}}
\usepackage{amsmath}

\begin{document}

\title{The Effect of Transparency on Students' Perceptions of AI Graders}


\author*[1]{\fnm{Joslyn} \sur{Orgill}}\email{jorgill2@illinois.edu}

\author[2]{\fnm{Andra} \sur{Rice}}\email{andralrice@gmail.com}

\author[1]{\fnm{Max}\sur{Fowler}}\email{mfowler5@illinois.edu}

\author[2]{\fnm{Seth} \sur{Poulsen}}\email{seth.poulsen@usu.edu}



\affil*[1]{\orgdiv{Siebel School of Computing and Data Science}, \orgname{University of Illinois Urbana-Champaign}, \orgaddress{\street{201 N. Goodwin Ave}, \city{Urbana}, \postcode{61801}, \state{IL}, \country{USA}}}

\affil[2]{\orgdiv{School of Computing}, \orgname{Utah State University}, \orgaddress{\street{4205 Old Main Hill}, \city{Logan}, \postcode{84322}, \state{UT}, \country{USA}}}


\abstract{The development of effective autograders is key for scaling assessment and feedback.
While NLP based autograding systems for open-ended response questions have been found to be beneficial for providing immediate feedback, autograders are not always liked, understood, or trusted by students. Our research tested the effect of transparency on students' attitudes towards autograders. Transparent autograders increased students' perceptions of autograder accuracy and willingness to discuss autograders in survey comments, but did not improve other related attitudes---such as willingness to be graded by them on a test---relative to the control without transparency. However, this lack of impact may be due to higher measured student trust towards autograders in this study than in prior work in the field. We briefly discuss possible reasons for this trend.}




\maketitle

\newcommand{\todo}[1]{\textcolor{red}{#1}}

\section{Introduction}

The use of generative models has expanded significantly in the field of education in the last few years~\citep{prather2025beyond,franklin2025generative,filippi2024large,yan2024practical}. This is due in part to the potential that generative models hold for implementation of artificially intelligent autograders, potentially decreasing the time spent grading and thus the time students wait for feedback. In the past few years, however, studies have shown that students are not keen on the use of autograders in assignments where their grade is affected~\citep{hsu_attitudes_2021,li2023wrong,rodway2023impact}. As students become more used to generative artificial intelligence (Gen AI) in the classroom and autograders in some courses \citep{prather2023robots}, it is possible that their attitudes may be changing. 

One source of contention for students is whether AI autograders can be trusted. Students may hold false folk theories about how these autograders work~\citep{hsu_attitudes_2021} and others may have concerns with respect to ethics~\citep{acosta2024knowledge}. This research aims to determine whether more transparency around \textit{how} artificially intelligent autograders function may help students to appropriately trust the autograder. Helping students become more comfortable with autograded assignments may make their use more comfortable and valuable for students.

Short answer, natural language responses are one kind of exercise that AI autograders have the opportunity to positively impact. Machine learning techniques have been a popular mechanism for short answering grading over the last decade~\citep{galhardi2018machine}. However, there are questions around the transparency, trustworthiness, and explainability of these graders~\citep{schneider2023towards,schlippe2022explainability}. One form of short answer question particularly relevant to CS are Explain in Plain English (EiPE) questions, the autograding of which has been met with skepticism and confusion from students in the past~\citep{li2023wrong,hsu_attitudes_2021}.

In this study, we seek to address students' misunderstandings and misgivings about AI autograders by providing them more information, and thus more transparency, about the autograders that give them feedback on their work. Specifically, we address the following research questions:
\begin{itemize}
\item[RQ1] 
To what degree does adding transparency about an autograder's accuracy increase student' perceptions of grader accuracy, helpfulness, and utility? 
\item[RQ2] How do students' experiences and views differ with respect to an AI autograder's transparency or lack thereof?
\end{itemize}

\section{Related Work}

\subsection{Student Perceptions of AI Graders}
There has been significant research into the development of high-accuracy systems for automatic short answer grading~\citep{bonthu2021automated,zhao2025language}. In comparison, there has been relatively little research on how to deploy such graders in real classrooms in ways that work well for student needs~\citep{zhao_autograding_2025,hsu_attitudes_2021,li2023wrong,kerslake_exploring_2025}.

\citet{zhao_autograding_2025} evaluated student performance and perceptions while using a natural language processing (NLP) autograder for proof by induction problems. The proofs were graded through an OpenAI GPT model, which is a type of NLP model. This model was not used to write the feedback. The immediate feedback provided was focused on the part of the rubric the student was missing or struggling with. Since the autograder model could pinpoint where the student had an issue, the appropriate pre-written feedback could be given to the student.
The results showed that the large language model (LLM) had comparable accuracy to human graders, and that students who used the autograder were able to write proofs with scores that were 11\% better than students who did not have access to the autograder. In spite of this, only about 50\% of the students said that they were satisfied with the grader and that it helped them improve their proofs. 


In another study, \citet{li2023wrong}
studied the effect of autograder errors on students. As autograder errors are unavoidable, it is necessary to understand what effects these errors have on students. This study researched the effects of autograder errors on student learning by inducing false positives and false negatives. They found that false positives harmed learning, as students were less likely to see the grader as committing an error when the grader graded the student as having submitted a correct response.
In contrast, false negatives were found to only harm learning for some participants. Other participants, who were likely more engaged in the material and activity, were not as affected by false negative autograder errors.


Perhaps most relevant to the current work is the study by
\citet{hsu_attitudes_2021}
on student attitudes towards and knowledge of AI autograders. The researchers investigated and discovered that students had inaccurate ``folk theories'' of autograders, which often led to poor answering strategies. Some of these folk theories included incorrect beliefs that the autograder only looked for keywords and that the autograders were developed without real student data.
Students also thought the autograder marked more things wrong than a human grader would, leading them to feel less satisfied with the grader. Furthermore, they wished they had received additional instruction on how to work with the autograder. The study ended with several suggestions on how to successfully use autograders, which we implemented in our own research to see if the recommendations will successfully mitigate the concerns.


\subsection{Explain in Plain English Questions}
\begin{figure}[h]
    \centering
    \includegraphics[width=1\columnwidth]{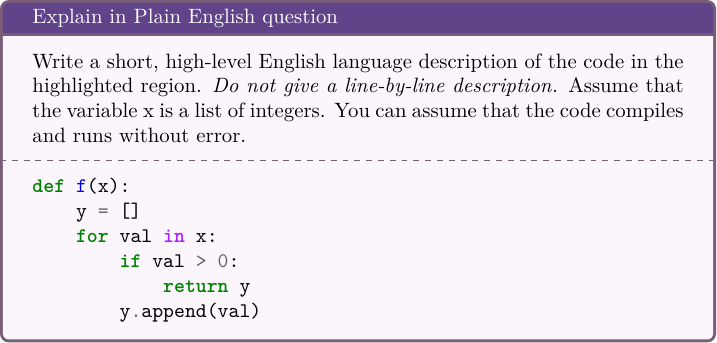}
    \caption{A sample EiPE question prompt}
    \label{fig:eipe_q}
\end{figure}

Learning to program has several composite skills, such as tracing, reading, and writing code~\citep{xie:theory-of-instruction:CSE:2019,lister2009further,lopez2008relationships,fowler2022reevaluating,newar2025mining}. Explain in Plain English (EiPE) questions are used to evaluate students' comprehension of code, as well as their ability to explain that comprehension~\citep{murphy2012explain,fowler2021should}. Typically, these questions take a form similar to Figure~\ref{fig:eipe_q}, presenting students with a piece of code and asking for a ``high-level'' description of that code. 

One barrier to the use of EiPE questions is the time required to assess them~\citep{fowler2021should}. Several efforts have been attempted to enable automatic grading and feedback, thus allowing for use of these questions at scale. \citet{fowler2021autograding} used a fairly simple logistic regression on a bag-of-words representation of real student answers. The model performed roughly as well as typical human graders, but it did not provide feedback beyond binary correctness. Later work compared this simple logistic regression approach to different natural-language processing approaches and large language model (LLM) based approaches. The LLM based approaches included using LLMs to generate code, an approach developed by \citet{smith2024code}, and using OpenAI GPT as a grader itself with few-shot examples of student work. At the time, GPT-4 as a grader had 86\% accuracy and the best graders were all in a similar range as the original logistic regression grader of 86\% to 88\%~\citep{fowler2024strategies}.

Another domain of AI-based grading used for EiPE is Code Generation Based Grading, using LLMs to generate code from student answers and then grading the questions via test cases~\citep{smith2024code}. This approach has expanded in a myriad of ways, including focusing on prompting as a skill~\citep{kerslake2024integrating,smith2024prompting,kerslake2025exploring} and leveraging the abilities of LLMs to work in multiple human languages so students can explain code in the language of their choice~\citep{smith2024eipl,prather2025breaking}. Given the effort needed to manually grade these items, effective autograding is useful to explore to enable the use of these questions at scale both formatively and summatively. 

EiPE questions have been recognized as being underutilized, but also a beneficial way for students to learn programming \citep{fowler2021should}. Studies like ours can help move the community towards more use of EiPE questions.

\subsection{Transparency in AI}
As this paper discusses findings surrounding the effects of transparency on student perceptions, it is necessary to define transparency in AI. Transparency in AI is often called for, yet it does not have one clear definition \citep{felzmann2020towards}. Research has called for transparency when using AI in educational contexts, including letting students know who had access to their data and what the data would be used for \citep{slade2013learning}. Other concerns around transparency are proposing that students know what algorithms are being used so they can determine if they are being graded or judged fairly \citep{ungerer2022ethical}. In this paper, we will discuss transparency as defined by \citet{memarian2023fairness}, which is concerned with providing a view into the ``black-box model'' used in grading. In providing transparency to the students, we are working to give them more insight into the models used. In particular, we provide students transparency by sharing the historical accuracy of the autograders and the size of the dataset they were trained on.

\subsection{Human-AI Collaboration in Education}
An emerging area of research is that on Human-AI collaboration \citep{fragiadakis2025evaluating,raees2024from}. Autograding is not often viewed as a human-AI collaboration, but it is easy to see how students collaborate with AI to learn more. As with any tool, there are more and less effective ways to use AI systems in education. 
Outside of education, transparency in AI systems has been shown to improve the experience for humans \citep{fragiadakis2025evaluating,kim2023help,raees2024from}.
We hypothesize that one of the problems with student AI grading and feedback systems is that when an AI system is used as a grader, students develop an adversarial relationship with the grader, rather than a collaborative one.
Viewing automated feedback and grading systems instead as a human-AI collaboration could open the door for students to have better experiences in having AI feedback to help them learn, while also being aware of the pitfalls.



\section{Methods}
To answer our research questions, we conducted a randomized controlled trial in which students completed a short learning activity, then afterwards responded to a survey about their experiences.
Students in the control condition were graded by the EiPE autograder as usual and did not receive any additional information. The students in the experimental condition received additional information about the autograding systems that graded their learning activities. 

\subsection{Study Population and Recruitment}
Participants for this study were recruited from an introductory computer science course for non-major students at a large public university in the United States.
The students were familiar with being asked EiPE questions in their class, with a bigram-based grader (similar to that described by~\cite{fowler2021autograding}) for EiPE questions.
Students were recruited during the second-to-last week of the semester, once they had learned most of the course content, to complete an additional learning activity for a small amount of extra credit. The students were randomly assigned to the experimental or control condition. There were 81 students in the treatment group and 74 students in the control.


\subsection{AI Grader}

The Explain in Plain English (EiPE) questions used in the study were graded by an autograder powered by OpenAI's GPT-4. This autograder used in this work is modeled after the few-shot LLM-based autograder described by~\citet{fowler2024strategies} and is not the same grader students had previously used in the course. While the autograder used regularly in class only gave students a right or wrong answer, the autograder used here gave students text-based feedback on how to improve. While students had familiarity with the bigram autograder, the one used in this study was quite different for these students.

A large language model was chosen because of its ability to provide immediate written feedback to students. The model was given specific instructions to ensure that the autograder would not immediately supply the student with the correct answer. 

To ensure the autograder gave appropriate feedback, the autograder was first prompted and then validated on real student data.
The model was prompted to act as a Computer Science teacher, providing the students a specific EiPE question that had been written several years before. The model received correct and incorrect example answers for that specific question. After this set up, the models were validated with real student responses. This is where the accuracy scores given to the students in the experimental condition were obtained. 

We chose the questions from a pool of EiPE questions that the autograder had previously been trained to grade. Our aim was to keep the accuracy scores above an 80\% threshold and to validate the model on at least 20 student responses.

\subsection{Experimental Materials}

The experiment consisted of the following parts:
\begin{enumerate}
    \item Consent form
    \item Pre-test (manually graded EiPE questions posthoc)
    \item Transparency statement
    \item Transparency quiz
    \item Learning Activity questions (autograded EiPE questions)
    \item Post-test (isomorphs of the pre-test, manually graded EiPE questions posthoc)
    \item Survey
\end{enumerate}

Students in the experimental condition were asked to read a transparency statement and quizzed on the content of the statement. The statement is shown in 
Figure~\ref{fig:autograder-explanation}. We note that the points in the statement were specific to this study and not the full set of recommendations that would be used on, say, a homework assignment (e.g., where students would be encouraged to seek teaching assistant support during the answer rewording process).

\begin{figure}
\begin{mdframed}
Please read:
In order to provide quicker feedback, we employ the use of \textit{language model autograders}.

Here is some information about the autograders we would like you to know before you get started:

\begin{itemize}
    \item The graders were developed using prior student data.
    \item This student data was validated by course staff and researchers with over a decade of combined experience with code reading exercises.
    \item Because they use real student data, the autograders have been calibrated to grading realistic student responses.
    \item The autograders achieve similar accuracy to human graders.
    \item The grader works based on the overall meaning of your response and does not look for specific key words or phrases.
    \item Because the autograders are occasionally wrong, feel free to simply reword your answer if you are graded as incorrect.
    \item If your answer is still graded as incorrect after rewording, consider making more substantial changes to your answer. ''
\end{itemize}
\end{mdframed}
\caption{Expository text shown to students in the experimental condition to address misconceptions about and help them to understand the autograders. After this text, students were asked questions like that in Figure~\ref{fig:example-comprehension} to ensure they paid attention to it}
    \label{fig:autograder-explanation}
\end{figure}

\begin{figure}
    \centering
    \includegraphics[width=\columnwidth]{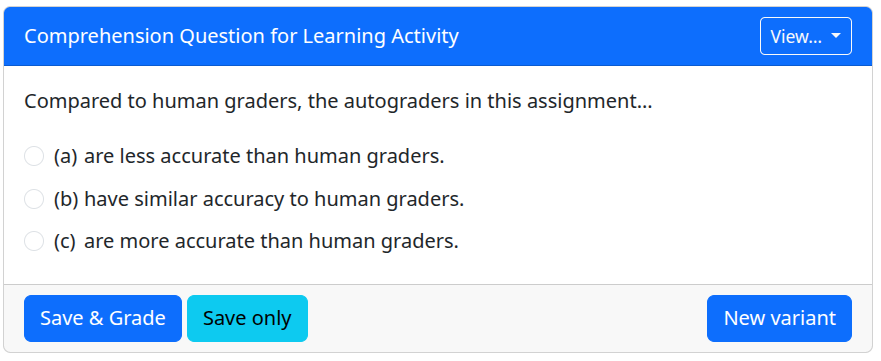}
    \caption{Example comprehension question asked of students in the experimental condition after reading the text in Figure~\ref{fig:autograder-explanation}. They were also asked true/false questions: ``A language model autograder is used to grade your responses.'', ``The graders in this assignment have been validated using real student data and comparing to human grader labels.'', and ``The grader is looking for specific words and not the overall meaning of your submission.'' }
    \label{fig:example-comprehension}
\end{figure}

The questions in the transparency quiz tested students' comprehension of whether the autograder grades based on the answer as a whole or on whether specific words were included, if real student data was used in the validation, and how the autograders' accuracy compared to human graders (see Figure~\ref{fig:example-comprehension}). The students were then given multiple attempts to answer the questions with the purpose of confirming their understanding of the autograder grading schemes.

\begin{figure}
    \centering
    \includegraphics[width=\columnwidth]{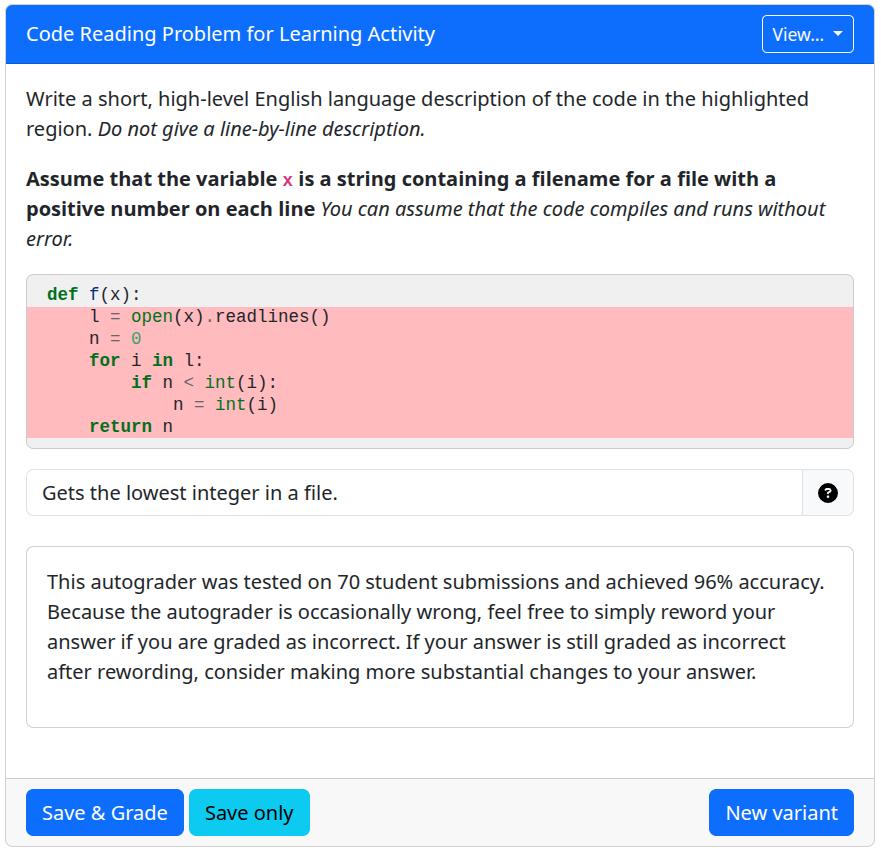}
    \caption{An example question shown to students in the experimental condition, including the transparency message.}
    \label{fig:example-problem_transparent}
\end{figure}

\begin{figure}
    \centering
    \includegraphics[width=\columnwidth]{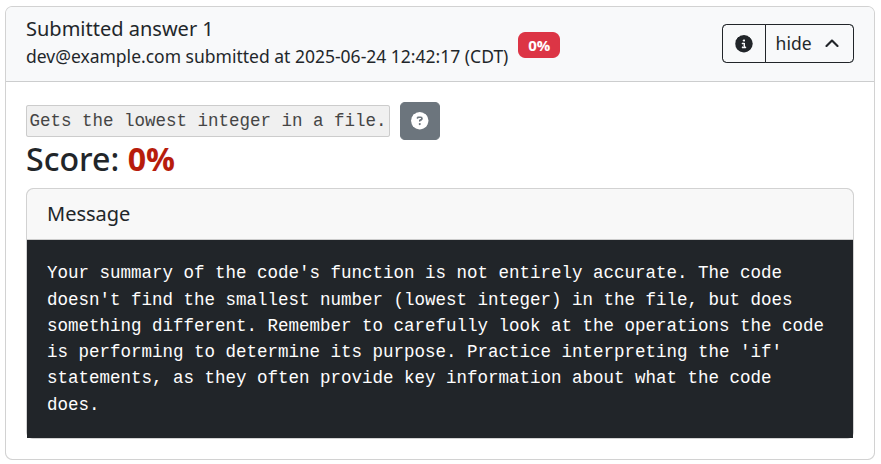}
    \caption{Example feedback generated by the AI autograder. The same grader and feedback mechanism were used for the experimental and control conditions.}
    \label{fig:example-problem-feedback}
\end{figure}

The learning activity questions were accompanied by an additional transparency statement. They each followed the format provided in Figure~\ref{fig:example-problem_transparent}, and gave feedback as shown in Figure~\ref{fig:example-problem-feedback}. Specifically, the transparency statement provided students with the number of real student submissions the grader was tested on and the accuracy the grader achieved. These numbers varied for each EiPE exercise in the set.


The control group was only told that the learning activity made use of an AI autograder. No transparency quiz was given and their questions did not contain extra information on the performance of each autograder. The EiPE exercises were otherwise the same.

\subsection{Post Survey}

After the post-test, we surveyed the student's attitudes on their experience with the autograder. The scores of each question were adapted from the Likert scale with agreement scores ranging from -2 to 2 with -2 being `Disagree', 0 being `Neutral', and 2 being `Agree'. To compare the responses to these Likert items between groups, we ran Mann-Whitney U tests, as the distributions of responses were non-normal. The set of questions analyzed from the survey is provided in Table~\ref{tab:mannwhitney_u_results}. The survey included questions on satisfaction with course Teaching Assistants that we omitted (originally, questions 1 and 7).





\subsection{Scoring Pre and Post-Test Answers}

The second author scored all the pre and post-test EiPE answers. The second author has several years of experience using EiPE questions in courses and conducting research on EiPE answers, making them well suited to marking the EiPE responses. Answers were scored on binary correctness, either a 0 (incorrect) or  1 (correct), using the same grading standards as found in previous EiPE autograder work. Answers were correct if they were technically accurate, written at a high level of abstraction, and did not include low-level details~\citep{fowler2021autograding}. As our primary interest for this study was the impact of \textit{transparency} on students' perceptions and attitudes, and not the impact this short task would have on student performance, we were not concerned with reconciling these grades over multiple graders.

\section{Results}

\subsection{Pre- and Post-Test: No Significant Difference in Performance Between Test Nor Group}

Before comparing the scores on the pre- and post-tests, responses from students who used outside resources on the activity were removed. This left 70 (out of 81) and 66 (out of 74) treatment and control students whose pre and post-test scores are valid. 

\begin{figure}
    \centering
    \includegraphics[width=1\columnwidth]{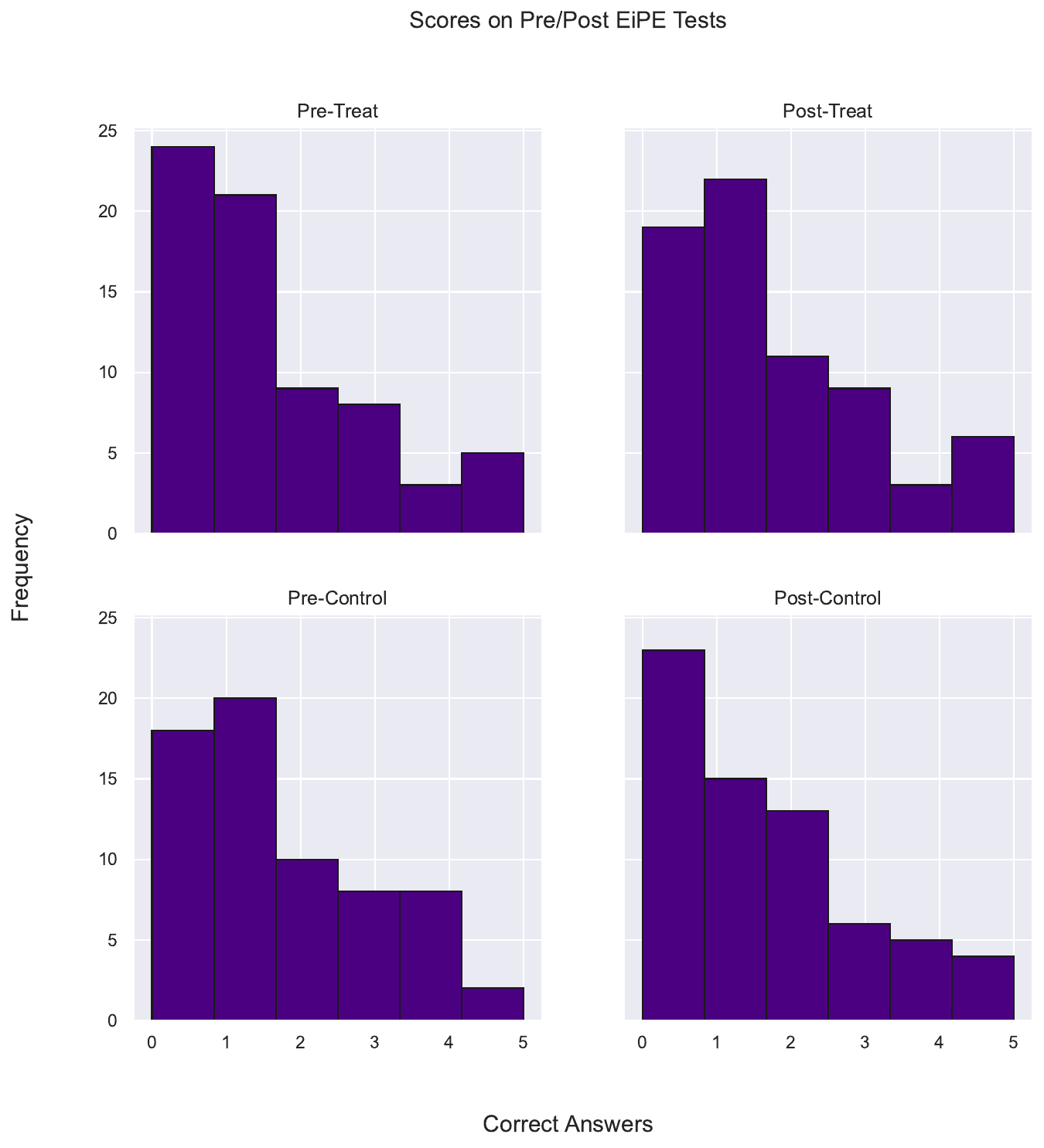}
    \caption{Performance on the pre and post-tests for EiPE questions. Scores were out of 5 for the number of questions answered correctly. There was no significant difference observed. 
    }
    \label{fig:test_distrib}
\end{figure}

The distribution of scores on the pre and post-test (0 for no points, 1 point for each question correct with a maximum of 5 for getting all questions correct) is provided in Figure~\ref{fig:test_distrib}. Running a Kruskal-Wallis Test found no significant difference between the pre and post-test scores, regardless of whether the student was in the control or treatment group ($H=1.11, p = 0.76$). Given the small number of questions on the pre and post-test and the small overall time dedicated to the experimental task, it is not too surprising there was not a significant improvement in performance. We graded the tests primarily to ensure that the two groups of students were roughly comparable.

\subsection{Survey Results}

\begin{table*}[h!]
    \centering
\caption{Mann-Whitney U test results between treatment and control. The treatment group found the grading statistically significantly more accurate ($p = 0.027$). The treatment group's positive perception of the autograder's helpfulness was close to significant ($p = 0.083$).}
\label{tab:mannwhitney_u_results}
    \begin{tabular}{|p{5.5cm}|x{1.7cm}|x{1.7cm}|x{1.4cm}|c|}
    \hline
    \textbf{Question} & 
    \textbf{Treatment Mean (s.d.)} & 
    \textbf{Control Mean (s.d.)} & 
    \textbf{Mann-Whitney U} & 
    \textbf{$p$} \\
    \hline
    Q2: I received accurate grading from this assignment's autograders. & 0.90 (0.82) & 0.59 (0.91) & 3568.5 & 0.03 \\
    Q3: I find the feedback from Teaching Assistants helpful in improving the English code explanations I am writing. &0.81 (0.82) &0.80 (0.78)& 3052.0 & 0.83 \\
    Q4: I find the feedback from this assignment's autograders helpful in improving the English code explanations I am writing. &0.85 (0.94)&0.66 (0.80)& 3442.0 & 0.08 \\
    Q5: I like the course's usual code-reading autograder. &0.28 (1.05)&0.12 (0.98)& 3284.5 & 0.28 \\
    Q6: I like this assignment's autograders. &0.73 (0.95)&0.77 (0.97)& 2920.0 & 0.77 \\
    Q8: Overall, I am satisfied with my experience of using this assignment's autograders. &0.89 (0.72)&0.78 (0.93)& 3123.5 & 0.61 \\
    Q9: I would like to have this assignment's autograders integrated in my course to give me feedback as I prepare for my exams. &1.07 (0.86)&0.95 (0.92)& 3232.0 & 0.37 \\
    Q10: I would like to have this assignment's autograders integrated in my course to grade my exams. &0.78 (1.07)&0.84 (1.02)& 2929.5 & 0.80 \\
    Q11: I would rather wait for a week to get human grading feedback than use this autograder for instant feedback. &-0.02 (1.21)&-0.05 (1.27)& 3006.0 & 0.98 \\
    \hline
    \end{tabular}
\end{table*}




\begin{figure*}
    \centering
    \includegraphics[width=1\textwidth]{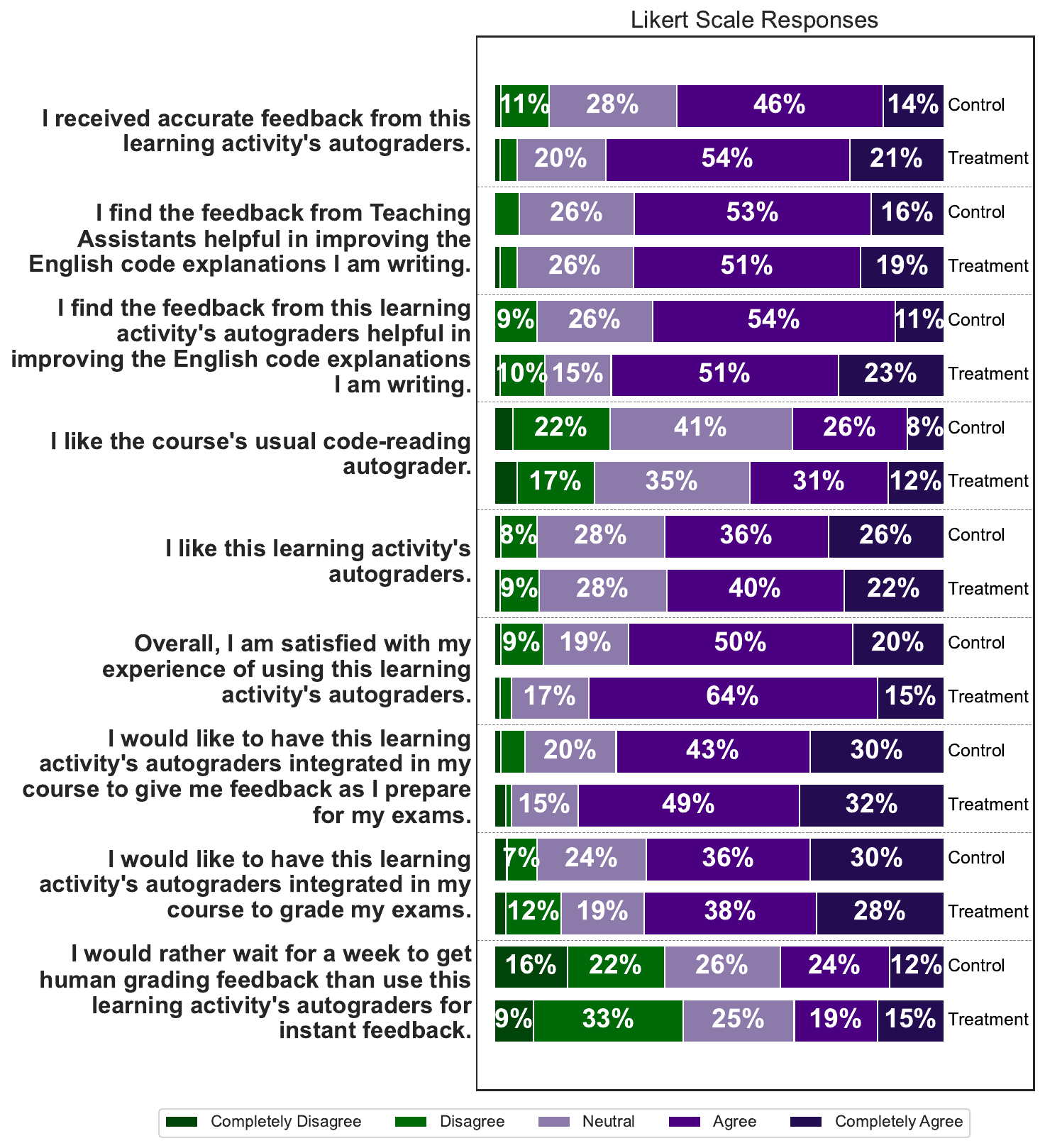}
    \caption{Survey for Control versus Treatment groups. The treatment group found the AI autograders to be more accurate and possibly more helpful than the control group, but the responses to other questions were similar across groups.}
    \label{fig:likert_quick}
\end{figure*}

There was a significant difference ($p = 0.027$) between the control and treatment group when comparing the perceptions of the autograders grading accuracy. The treatment group on average ($\mu = 0.90, \sigma = 0.82$) rated the accuracy of the autograder to be far higher than the control group did on average ($\mu = 0.59, \sigma = 0.91$).

Additionally, there was a marginally significant difference ($p=0.083$) in the perception of how helpful the autograder was in improving the student's plain English code explanations. Again, the treatment group ($\mu = 0.85, \sigma = 0.94$) found the autograder more helpful on average than the control group ($\mu = 0.66, \sigma = 0.80)$ on average found the autograder.

Regardless of experimental group, the students indicated a preference for immediate feedback through an autograder in contrast to delayed feedback from a human grader (see Table \ref{tab:mannwhitney_u_results}). 
The students also indicated a willingness to use the autograder for preparing for and grading of exams. The latter was less desired by the students (see Figure \ref{fig:likert_quick}).

\subsection{Thematic Analysis of Survey Questions}
The survey also included four open ended questions that asked students for positive and negatives experiences with the autograder, as well as how it could be useful and any additional comments. Two authors conducted a thematic analysis of these responses. The authors coded all students' answers to the four questions in three rounds: in two sets of ten answers each to discuss initial codes and calibrate understanding, then coding the rest of the answer set. In a final meeting, the authors met to discuss emerging themes. In the rare case of disagreement (only six instances of disagreement), the disagreements were discussed and consensus reached. The different themes and counts can be found in Tables \ref{tab:positive_experience}--\ref{tab:other_comments}. This section will spend time discussing the broad themes of this qualitative analysis. Students are identified with a number and whether they were in the treatment or control condition. For example, T29 is the 29th student from the treatment (transparent autograder) group. Words in brackets are added for clarity to the quotes.

\subsubsection{Positive Themes}
\begin{table}[tb]
\centering
\caption{Count and percentage of different themes for the question ``Can you share a positive experience you had with the autograder?'' (percentage calculated by number of respondents). Dashes indicate no findings for that group and theme.}
\begin{tabular}{ccccc}
          & \multicolumn{2}{c}{Control (\textit{n} = 44)} & \multicolumn{2}{c}{Treatment (\textit{n} = 53)} \\ \midrule
          & Count    & Percentage & Count   & Percentage   \\ \midrule
Personalized Feedback & 21 & 47.73\% & 24 & 45.28\% \\
Good for Learning & 20 & 45.45\% & 19 & 35.85\%  \\
Useful & 9 & 20.45\% & 11 & 20.75\%  \\
Speed of Feedback & 6 & 13.64\% & 11 & 20.75\%  \\
Gives Answer Away & 1 & 2.27\% & - & -  \\
Doesn't Give Answer Away & - & - & 2 & 3.77\%  \\
More Points & 1 & 2.27\% & 2 & 3.77\% \\
Accurate & 1 & 2.27\% & 2 & 3.77\% \\
Satisfying & 1 & 2.27\% & 4 & 7.55\% \\
Too Lenient & 1 & 2.27\% & - & - \\ \midrule
Mean Number of Characters & \multicolumn{2}{c}{57.20} & \multicolumn{2}{c}{68.62} \\
Standard Deviation & \multicolumn{2}{c}{55.13} & \multicolumn{2}{c}{47.36}
\end{tabular}
\label{tab:positive_experience}
\end{table}

Common themes in the open-ended questions were about the personalized and speedy feedback that the autograder gave. One student said, \textit{``getting immediate, specific feedback helped me fix the answer to the question''} (T48) as a positive experience with the autograder. The personalization of the autograder was also a positive theme, like one student mentioning that the autograder \textit{``It [the grader] allows you to access what you need to fix''} (T23).

A similar theme was the effect that autograders had on learning. Many students mentioned that autograders were good for learning, with one student commenting, \textit{``I like when it gives other ways to answer the same question, it helps me understand what the code is saying''} (T41). Students seem to believe the autograder can help them learn the concepts more effectively.

Of note to the research team was the students appreciation of the convenience of the autograder. One student mentioned \textit{``Autograders are very helpful when I'm studying late and don't have anyone to respond till to later''} (T44). Autograders are perhaps more able to fit in with the busy schedule of a student, as well as provide on-demand feedback at hours where other support is not immediate, e.g., late at night.

Students also described the confidence that the autograder gave them as a useful byproduct of using the autograder (\textit{``I have a better understanding of what I am doing wrong and how I should approach code-description questions without feeling so afraid and clueless''} (C5)). 

 Overall, there was a clear signal of student positivity for the autograder and the autograder's most positive traits, such as personalization and speediness. Interestingly, while students in the control group had an mean of 57.20 characters per response for the question about positive experiences, the students in the treatment group had an mean of 68.62 characters, suggesting that students in the treatment group had more positive experiences to share, or at least more to say about them (see Table \ref{tab:positive_experience}). While this difference between the groups is not statistically significant, it is interesting that the control group had more negative things to say than the treatment group. These findings would seem to indicate that when students know more about the autograder through transparent interventions, students tend to comment more about the autograder and they are kinder in their comments.

\subsubsection{Negative Themes}
\begin{table}[tb]
\centering
\caption{Count and percentage of different themes for the question ``Can you share a negative experience you had with the autograder?'' (percentage calculated by number of respondents). Dashes indicate no findings for that group and theme.}
\begin{tabular}{ccccc}
          & \multicolumn{2}{c}{Control (\textit{n} = 44)} & \multicolumn{2}{c}{Treatment (\textit{n} = 53)} \\ \midrule
          & Count    & Percentage & Count   & Percentage   \\ \midrule
None & 8 & 18.18\% & 12 & 22.64\% \\
Too Strict & 13 & 29.55\% & 14 & 26.42\%  \\
Disagree and Confusing & - & - & 3 & 5.66\%  \\
Unclear & 6 & 13.64\% & 10 & 18.87\%  \\
Disagree with Feedback & 7 & 15.91\% & 6 & 11.32\%  \\
Contradictory & 3 & 6.82\% & 2 & 3.77\%  \\
Bad Feedback Loop & 1 & 2.27\% & 6 & 11.32\%  \\
Too Slow & 3 & 6.82\% & - & - \\
Gives Away Answer & 1 & 2.27\% & - & - \\
Buggy & 2 & 4.55\% & 1 & 1.89\% \\
Hard Questions & 2 & 4.55\% & 3 & 5.66\% \\
No Feedback & 3 & 6.82\% & - & - \\
Frustrating & 2 & 4.55\% & 1 & 1.89\% \\ \midrule
Mean Number of Characters & \multicolumn{2}{c}{65.14} & \multicolumn{2}{c}{65.98} \\
Standard Deviation & \multicolumn{2}{c}{71.43} & \multicolumn{2}{c}{66.37}
\end{tabular}
\label{tab:negative_experience}
\end{table}
A common theme in the question about negative experiences in both treatment and control groups was the strictness of the model. Students mentioned that even if they perceived the issue with their answer to be one word being spelled wrong or omitted, they were getting the question marked as wrong overall. Comments suggested frustration with this aspect of the autograder (\textit{``I was getting marked off for missing one word in my answer''} (C9)). Similar proportions of students indicated this in both the treatment and the control groups, suggesting that even with transparency, students still felt that the autograder was too strict. As discussed in \citet{fowler2021should}, the strictness by which EiPE questions are graded can be adjusted from the ground truth based on the desires of the instructor. The comments about strictness are something that instructors who choose to use autograders can decide whether to adjust the desired strictness or not.

Another group felt that the responses from the autograder weren't clear (\textit{``It was unclear on what I had to fix about my solution''} (T10)). This was an often-seen theme that may suggest refining that can be done on the autograder.

Overall, there were few comments in the responses that suggested students hated the autograder, although there were some who did. Most students saw the benefits of an autograder like the one in the study, and many students simply did not have any negative experiences they wanted to share in the comments.

\begin{table}[tb]
\centering
\caption{Count and percentage of different themes for the question ``In your opinion, how can the autograder be useful in your learning experiences?'' (percentage calculated by number of respondents). Dashes indicate no findings for that group and theme.}
\begin{tabular}{ccccc}
          & \multicolumn{2}{c}{Control (\textit{n} = 44)} & \multicolumn{2}{c}{Treatment (\textit{n} = 53)} \\ \midrule
          & Count    & Percentage & Count   & Percentage   \\ \midrule
Good for Learning & 14 & 31.82\% & 21 & 39.62\% \\
Speed of Feedback & 13 & 29.55\% & 20 & 37.74\% \\
Personalized Feedback & 18 & 40.91\% & 28 & 52.83\% \\
No Fear & 1 & 2.27\% & 1 & 1.89\% \\
Convenient & 2 & 4.55\% & 1 & 1.89\% \\
Unclear & 1 & 2.27\% & - & - \\
Good Practice & 2 & 4.55\% & 4 & 7.55\% \\
EiPE is Hard & 1 & 2.27\% & - & - \\
Bad Feedback Loop & - & - & 1 & 1.89\% \\
Beware Over-reliance & - & - & 1 & 1.89\% \\
Grow Knowledge & - & - & 3 & 5.66\% \\
Useless & - & - & 1 & 1.89\% \\ \midrule
Mean Number of Characters & \multicolumn{2}{c}{58.05} & \multicolumn{2}{c}{78.66} \\
Standard Deviation & \multicolumn{2}{c}{40.43} & \multicolumn{2}{c}{76.68}
\end{tabular}
\label{tab:useful_to_learning}
\end{table}

\begin{table}[tb]
\centering
\caption{Count and percentage of different themes for the question ``Do you have other comments on the autograder?'' (percentage calculated by number of respondents). Dashes indicate no findings for that group and theme.}
\begin{tabular}{ccccc}
          & \multicolumn{2}{c}{Control (\textit{n} = 44)} & \multicolumn{2}{c}{Treatment (\textit{n} = 53)} \\ \midrule
          & Count    & Percentage & Count   & Percentage   \\ \midrule
None & 22 & 50.00\% & 29 & 54.72\% \\
Overall Positive & 6 & 13.64\% & 9 & 16.98\% \\
Overall Negative & - & - & 1 & 1.89\% \\
Speedy Feedback & 1 & 2.27\% & 2 & 3.77\% \\
AI Mistrust & 2 & 4.55\% & 1 & 1.89\% \\
Needs Verification & 2 & 4.55\% & - & - \\
Room to Grow & 2 & 4.55\% & - & - \\
Flexible & - & - & 2 & 3.77\% \\
Grading Concerns & - & - & 2 & 3.77\% \\
Tricky & - & - & 1 & 1.89\% \\
More Explanation & - & - & 1 & 1.89\% \\
Mitigate Fear with Policy & - & - & 1 & 1.89\% \\ \midrule
Mean Number of Characters & \multicolumn{2}{c}{13.16} & \multicolumn{2}{c}{34.26} \\
Standard Deviation & \multicolumn{2}{c}{27.77} & \multicolumn{2}{c}{64.19}
\end{tabular}
\label{tab:other_comments}
\end{table}

\subsubsection{Trust in AI Graders}
An interesting theme that emerged from the student answers was the way that students trusted the autograder. Some students seemed inclined to trust the autograder only to a certain point. Students specified that for practice, the autograder was fine, but not for real assignments. 

One interesting comment represented the view of this theme: \textit{``It can be useful for providing feedback, but should not be relied upon''} (T19). Similar comments mentioned that the grading should be verified by a human grader. These themes seem to indicate a mistrust when the autograder is relied on too heavily by teachers. Students may be more content when the autograder is used for practice, but when used for actual grades, they prefer having a human grader validate the grades to guarantee correct grading. Notably, in the context of the course this study was conducted in, student answers to these kinds of questions using a different type of natural-language processing autograder were validated by course staff during exams. Familiarity with this practice may have driven its appearance in student commentary to some degree.

\subsubsection{Willingness to Share Opinions}
An unexpected finding from the analysis of the open-ended questions was the willingness of students in the treatment condition to comment more than students in the control condition. For three of the questions, the average number of characters in the answer was much more in the students in the treatment group than the students in the control group. The final question about comments was statistically significant according to a Mann-Whitney $U-$test comparing the count of characters ($p=0.043$). Interestingly, the only question where the number of characters used was almost the same was the question asking students to share negative experiences with the autograder.

From this analysis, it would suggest that by giving students more transparency about the autograder, students were more likely to want to comment on the autograder. These results are obviously preliminary, but they do suggest an interesting pattern for future study.

\section{Discussion}

Throughout the course of the research, a few factors differed from previous studies on student attitudes towards autograders. First, we compared the addition of a transparency statement on student attitudes. With the inclusion of this statement, students in the treatment group found the autograder significantly more accurate than the control group. 
Looking into past studies~\citep{hsu_attitudes_2021,li2023wrong},
we note their results suggested students 
tended to have a more negative view on autograder accuracy.
Explaining how the autograder works and its past accuracy may help students better understand the level of accuracy the autograder provides. In comparing how helpful the autograder was for the learning activity, the treatment group reported the autograder as, in the case of statistical significance, marginally more helpful ($p=0.083$) than what the control group reported for the autograder.

While prior work reported students having some skepticism towards autograders ~\citep{hsu_attitudes_2021,zhao_autograding_2025},
our results indicate what may be a slight upward trend in students' interest in using autograders for exam preparation, along with higher satisfaction with and trust in the autograder, regardless of experimental condition.  
There were some key experimental setup differences that may have influenced this development. First, the experiments for both prior studies by \citet{hsu_attitudes_2021} and \citet{zhao_autograding_2025}
took place at least a year or more prior to when our experiment took place. This introduces an element of time during which students may have grown accustomed to the use of AI in a classroom setting as LLMs became more prevalent. Further, the different style of the LLM powered autograder compared to the course's existing autograder may have meant the students preferred this new autograder due to novelty and the differences over the course's standard grader.

Despite the treatment and control group having different perceptions on the autograder's helpfulness and accuracy, these perceptions were not accompanied by a difference in their desires to use the autograder on homework or even on exams. 
Improved perceptions of autograder accuracy were thus not a prerequisite for wanting to use an autograder, even in high stakes settings. Again, this may be due to the course context. As students in the course already had autograded questions, and could already appeal perceived issues with autograder accuracy with the course staff, this familiarity with autograders and appeals processes may mean students are fine with autograders \textit{so long as} they can remediate negative impacts when necessary.

One alternative theory as to why our results differed from previous studies is that students may be less willing to use AI grading for high complexity tasks. We propose this theory in junction with the results reported by \citet{zhao_autograding_2025}. This previous study had students using an autograder for grading proof by induction problems, which is a more complex task than Explain in Plain English (EiPE) questions. The students in this past research indicated a fairly low trust in the feedback from the autograder relative to students in our research, who on average found the autograder to be helpful and accurate. Students who worked on proofs \citep{zhao_autograding_2025} notably rated human grading more positively than they rated the autograded which contrasts from our research where the students rated the human and autograders comparatively. The lower stakes and more straightforward nature of EiPE exercises may lend itself to students' trust more quickly than complicated tasks like proof grading.

\section{Conclusions and Future Work}
In this paper, we presented a randomized controlled trial which measured student perceptions of AI autograders under different conditions.
We showed that giving students additional information about the AI graders they were being assessed by, increased the students' perception of the grader's accuracy, and potentially their perceptions of the grader's helpfulness. However, it did not affect other measurements, such as their willingness to use the autograder on a homework or exam.

This work lays a foundation for future working in understanding the relationship between students and autograders. Future work can focus on additional ways to help increase student trust in autograders. Longitudinal studies, seeing if interventions of this sort have a bigger impact over the course of a semester rather than just in a one-hour activity, could also be useful. Future work could also address how student interactions with autograders affect their learning and knowledge retention, rather than just their perceptions of the tool.

\bibliography{references}

\end{document}